\documentstyle[12pt]{article}
\begin{document}
\begin{center}
\hbox to \hsize{\hfil hep-ph/9610241}
{\large\bf Spin and Flavor Content of Constituent Quarks and One-Spin
Asymmetries in Inclusive Processes\\}
\vspace{5mm}
{ \underline{S. M. Troshin}
 and N. E. Tyurin}\\
\vspace{5mm}
{\small \it Institute for High Energy Physics,\\
  Protvino, Moscow Region, 142284 Russia\\}
\end{center}
\begin{center}
\begin{minipage}{130mm}
\small
We discuss a nonperturbative mechanism  for one-spin
asymmetries in inclusive hadron production.
The main role in
generation of this asymmetry belongs to the orbital angular momentum
of  quark-antiquark cloud in  internal structure of
constituent quarks.
\end{minipage}
\end{center}
 As it is widely known now, only part (less
than one third in fact) of the proton spin is due to quark spins.
These results can be interpreted in the
effective QCD approach ascribing a substantial part of hadron spin
to an orbital angular momentum of quark matter.  It is natural to
guess that this orbital angular momentum might be revealed in
asymmetries in hadron production.

 We consider a nonperturbative hadron to be consisting of the
constituent quarks located at the core of the hadron and
quark condensate surrounding this core.
We refer to effective QCD and use the NJL model
[1] as a basis. The  Lagrangian  in addition to the
$4$--fermion interaction of the original NJL model includes
$6$--fermion $U(1)_A$--breaking term.

Strong interaction radius of this quark is determined by its Compton
wavelength:  $r_Q=\xi/m_Q$, where the constant $\xi$ is universal for
different flavors.  Spin of constituent quark $J_{U}$  in this
 approach is determined by the  sum
\[
J_{U}=1/2=J_{u_v}+J_{\{\bar q
 q\}}+\langle L_{\{\bar qq\}}\rangle .
\]   Value of the orbital
 momentum contribution into the spin of constituent quark can be
 estimated with account for the  experimental results from
 deep--inelastic scattering.  The important point  what the origin of
 this orbital angular momentum is. It was proposed [2] to
  consider an analogy with an  anisotropic generalization of the
 theory of superconductivity which seems to match well with the above
 picture for a constituent quark.

 We  consider the hadron processes of
 the  type \[ h_1^\uparrow +h_2\rightarrow h_3 +X\] with polarized
 beam or target and $h_3$ being a charged pion.

 In the model constituent quarks  are supposed to scatter in a
quasi-independent way by some effective field which is being
 generated at the first stage of interaction under overlapping of
peripheral condensate clouds [2].  Inclusive production
  of hadron $h_3$  results from recombination of the constituent
 quark (low $p_{\perp}$'s, soft interactions) or from the excitation
 of this constituent quark, its decay and subsequent fragmentation in
 the hadron $h_3$. The latter process is determined by the
 interactions at distances smaller than constituent quark radius and
is associated therefore with hard interactions (high $p_{\perp}$'s).
Thus, we adopt a two--component picture of hadron production which
 incorporates interactions at large and small distances.  We supposed
  that $\pi^+$--mesons are produced mostly by up flavors and
$\pi^-$--mesons --- by down flavors.

In the expression for asymmetry $A_N$ [2] \begin{equation}
A_N(s,\xi)=\frac{\int_0^\infty bdb I_-(s,b,\xi)/|1-iU(s,b)|^2}
{\int_0^\infty bdb I_+(s,b,\xi)/|1-iU(s,b)|^2} \label{xnn}
\end{equation} the function $U(s,b)$ is the generalized reaction
matrix [3] and the functions $I_{\pm}$ can be expressed
through its multiparticle analogs [2], $\xi$ denotes
the set of kinematical variables for the detected meson.
  In the model the
spin--independent part $I_+(s,b,\xi)$ gets contribution from the
processes at small (hard processes) as well as at large (soft
processes) distances, i.e.  $I_+(s,b,\xi)= I^h_+(s,b,\xi)+
I^s_+(s,b,\xi)$, while the spin--dependent part $I_-(s,b,\xi)$ gets
contribution from the interactions at short distances only
$I_-(s,b,\xi)=I^h_-(s,b,\xi)$.  The function $I_-^h(s,b,\xi)$ gets a
nonzero value due to  interference between the two helicity
amplitudes, which gain different phases due to internal motion of
partons inside the constituent quark.  The following relation between
the functions $I_-^h(s,b,\xi)$ and $I_+^h(s,b,\xi)$ has been proposed
assuming  the effect of internal motion of partons inside constituent
quark  leads to a shift in the produced meson transverse momentum:
\begin{equation} I_-^h(s,b,\xi)= \sin[{\cal{P}}_{\tilde Q}(x) \langle
L_{\{\bar q q\}}\rangle] I^h_+(s,b,\xi), \end{equation} where
${\cal{P}}_{\tilde Q}(x)$ is the polarization of the leading
constituent quark $\tilde Q$ (in the process  of the meson $h_3$
production) and $\langle L_{\{\bar q q\}}\rangle$ is the mean value
of internal angular momentum inside the constituent quark. This
relation allowed to get a parameter--free prediction for the
$p_{\perp}$--dependence of asymmetries in inclusive pion production
[2].

The $x$--dependencies of the functions $I_+^s(s,b,\xi)$ and
$I_+^h(s,b,\xi)$ are determined by the distribution of constituent
quarks in a hadron and by the structure function of constituent quark
respectively:  \begin{equation} I_+^s(s,b,\xi)  \propto
\omega_{\tilde Q/h_1}(x) \Phi^s(s,b,p_{\perp})\quad \mbox{and} \quad
I_+^h(s,b,\xi)\propto \omega_{\tilde q/\tilde Q}(x)
\Phi^h(s,b,p_{\perp}). \label{is} \end{equation}
Taking into account the above
relations, we can represent inclusive cross--section in the unpolarized
case
$d\sigma/d\xi$ and asymmetry  $A_N$ in the
following forms:
\begin{equation}
\frac{d\sigma}{d\xi}=8\pi
[W_+^s (s,\xi)+W_+^h(s,\xi)],\label{cs}
 \end{equation}

 \begin{equation} A_N(s,x,p_{\perp})=\frac{ \sin[{\cal{P}}_{\tilde Q}(x)
\langle L_{\{\bar q q\}}\rangle] {W_+^h(s,\xi)}}{ {
[W_+^s (s,\xi)+W_+^h(s,\xi)]}},\label{an}
 \end{equation}
where the  functions $
W_+^{s,h}$ are determined by the interactions at large and small
distances:  \[ W_+^{s,h}(s,\xi)=\int_0^\infty
bdb{I_+^{s,h}(s,b,\xi)}/ {|1-iU(s,b)|^2}.  \]
 The asymmetry in the model
  has a
significantly different $x$-dependence in the regions of transverse
 momenta $p_{\perp}\leq \Lambda_\chi$ and $p_{\perp}\geq
\Lambda_\chi$.
 We have simple
$p_{\perp}$--independent expression for asymmetry at
$p_{\perp}>\Lambda_\chi$ \begin{equation} A_N(s,x,p_{\perp})\simeq
 \sin[{\cal{P}}_{\tilde Q}(x) \langle L_{\{\bar q q\}}\rangle]
\label{las} \end{equation} and increasing asymmetry for
$p_{\perp}\leq\Lambda_\chi$ \begin{equation}
 A_N(s,x,p_{\perp})\simeq\sin[ {\cal{P}}_{\tilde Q}(x) \langle
L_{\{\bar q q\}}\rangle] \frac{\omega_{\tilde q/\tilde Q}(x)}
{\omega_{\tilde Q/h_1}(x)} r(s,p_{\perp}). \label{sas} \end{equation}

We consider the simplest possible dependence, e.g. the  linear
  one ${\cal{P}}_{\tilde Q}(x)=
 {\cal{P}}_{\tilde Q}^{max}x$,
 where
 ${\cal{P}}_U^{max}= -{\cal{P}}_D^{max}=1$.
 The  value of $\langle L_{\{\bar q q\}}\rangle\simeq
 0.4$ has been taken [2] on the basis
 of the DIS data.
 Using the above value
 of angular orbital momentum we obtain a good agreement with the data.
Note also that in this model the
 mirror symmetry for asymmetries in $\pi^+$ and $\pi_-$ production at
 $p_{\perp}>\Lambda_\chi$, i.e. $A_N^{\pi^+}=-A_N^{\pi^-}$ takes
 place.  Due to this relation asymmetry in $\pi^o$ production
 $A_N^{\pi^o}>0$ since from experimental data it is known that
$d\sigma^{\pi^+}/d\xi> d\sigma^{\pi^-}/d\xi$.

The predicted $p_{\perp}$-dependence of asymmetry has recently
 been confirmed by the experimental data [4].

The proposed model allows one to get the description of the spin effects
in the hyperon production also [5].

Authors  express their  gratitude to G. Goldstein for interesting
discussions and one of them (S.T.) to Organizing
Committee of SPIN96 Simposium for financial support.
\vspace{0.2cm}
\vfill
{\small\begin{description}
\item{[1]} Y.  Nambu and G.
Jona-Lasinio, Phys.  Rev. \bf 122 \rm  (1961) 345;\\   V.  Bernard, R.
L. Jaffe and U.--G.  Meissner, Nucl.  Phys. \bf B308 \rm  (1988) 753;\\
 K.Steininger and W.  Weise, Phys. Lett.  \bf
B329 \rm  (1994) 169.
\item{[2]} S. M. Troshin and N. E.  Tyurin,
Phys. Rev.  \bf D52 \rm  (1995) 3862;
 Phys. Lett.  \bf
 B355\rm (1995) 543;
Phys. Rev.  \bf D54 \rm  (1996) 838.
\item{[3]}  A. A. Logunov, V. I. Savrin,
N. E. Tyurin and O. A. Khrustalev, Teor.  Mat. Fiz.  \bf  6\rm (1971)  157.
\item{[4]} A. Bravar, Talk given at SPIN96 Simposium, Amsterdam, 10-14 September, 1996
(these Proceedings).
\item{[5]} S. M. Troshin and N. E.  Tyurin, Preprint IHEP 96-12, Protvino, 1996.
\end{description}}
\end{document}